\documentclass[a4paper,5p]{elsarticle}
\usepackage{amssymb}
\usepackage{amsmath}
\usepackage{graphicx}
\usepackage{xspace}
\usepackage{tabularx}
\usepackage{color}
\usepackage{cuted}
\usepackage[normalem]{ulem}

\newcommand{\add}[1]{\textcolor{black}{#1}} 

\usepackage{amsmath,amssymb,amsfonts,stmaryrd,wasysym,graphicx,multirow,textcomp,subfigure}
\usepackage{url}
\usepackage[colorlinks=true, citecolor=cyan, urlcolor=cyan]{hyperref}
\usepackage{romannum}
\usepackage{mathtools}
\usepackage[utf8]{inputenc}
\usepackage{braket}
\usepackage{xcolor}
\usepackage{color,soul} 
\usepackage{bm,xfrac}
\usepackage[normalem]{ulem}
\usepackage{enumitem}
\usepackage{dcolumn}
\usepackage{bm}
\usepackage[flushleft]{threeparttable}
\usepackage{array}
\usepackage{booktabs}
\usepackage{float}
\usepackage{setspace}
\usepackage{longtable}
\usepackage{balance}
\usepackage[version=4]{mhchem}

\begin{document}

\setcounter{page}{1}
\title{
Normal Dirac Semimetal Phase and  Zeeman-Induced Topological Fermi Arc in \ce{PtSr5}}

\author[1]{Inkyou Lee\fnref{equal}}
\author[1]{Churlhi Lyi\fnref{equal}}
\author[1]{Youngkuk Kim\corref{cor1}}
\ead{youngkuk@skku.edu}
\cortext[cor1]{Corresponding author}
\address[1]{Department of Physics, Sungkyunkwan University, Suwon 16419, Korea}
\date{\today}

\begin{abstract}
\add{Pt–Sr binary intermetallics encompass a broad range of stoichiometries and crystal structures, stabilized by complex bonding and multivalent chemistry. The Sr-rich end member, \ce{PtSr5}, is recently identified via artificial-intelligence–guided materials design as a body-centered tetragonal compound ($I4/m$)~\cite{Merchant23p80}.} Using first-principles calculations, we show that \ce{PtSr5} hosts a Dirac semimetal phase with trivial $\mathbb{Z}_2$ topology\add{, classified as a normal Dirac semimetal}. A symmetry-indicator analysis based on parity eigenvalues at the eight time-reversal-invariant momenta confirms that all $\mathbb{Z}_2$ invariants—evaluated on time-reversal-invariant two-dimensional subspaces of momentum space with a direct band gap—are trivial, thereby establishing the topologically trivial nature of the Dirac semimetal phase. Nonetheless, our calculations reveal that applying an external Zeeman magnetic field along the $z$-axis drives the system into a Weyl semimetal phase, as corroborated by characteristic changes in the computed surface states. This work demonstrates the tunability of topological phases in \ce{PtSr5} via external perturbations and highlights the effectiveness of AI-based materials exploration in discovering new quantum materials.
\end{abstract}


\maketitle







\section*{Introduction}

The discovery of topological materials \cite{Kane05p226801, Bernevig06p1757, Fu07p045302, Hasan10p3045, Qi11p1057, Armitage18p015001} has significantly advanced our understanding of condensed matter systems, offering new insights into fundamental quantum phenomena and paving the way for novel applications in electronics \cite{Xu11p186806}, spintronics \cite{Xu15p613}, and quantum computing \cite{Huang15p7373}. The identification of topological insulators \cite{Hasan10p3045, Qi11p1057}, characterized by insulating bulk states and robust conducting surface states protected by time-reversal symmetry, laid the foundation for this rapidly growing field. Following this, the realization of topological semimetals---materials \cite{Weng15p011029, Gao19p153} hosting band crossings that lead to exotic quasiparticles such as Weyl \cite{Xu15p613, Lv15p031013} and Dirac fermions \cite{Liu14p864, Liu14p677, Liu14p864, Neupane14p3786}--marked another major breakthrough. Weyl semimetals, featuring Weyl points with well-defined chirality, have been experimentally observed in materials such as TaAs \cite{Xu15p613} and MoTe$_2$ \cite{Deng16p1105}, where their unique Fermi arcs and anomalous transport properties have been extensively studied \cite{Huang15p7373, Soluyanov15p495}. Similarly, Dirac semimetals such as Na$_3$Bi \cite{Wang12p195320, Liu14p864} and Cd$_3$As$_2$ \cite{Wang13p125427} have provided a solid-state platform for studying relativistic Dirac fermions \cite{Vafek14p83}, deepening our understanding of quantum anomalies \cite{Nielsen83p389}, chiral transport \cite{Son13p104412}, and unconventional responses to external fields \cite{Xiong15p413}, thereby expanding the realm of both fundamental physics and applications.

\begin{figure}[t!]
    \centering
    \includegraphics[width=0.48\textwidth]{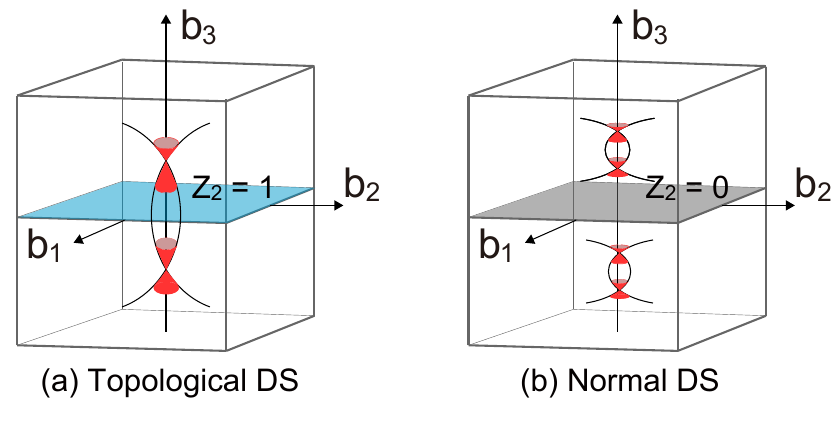}
    \caption{
    (a) Topological Dirac semimetal (DS) with a nontrivial $\mathbb{Z}_2$ topological invariant ($\mathbb{Z}_2=1$) defined on the $b_3=0$ sub-plane. Such a phase is realized in materials like Na$_3$Bi and Cd$_3$As$_2$. 
    (b) Normal Dirac semimetal with a trivial $\mathbb{Z}_2$ invariant ($\mathbb{Z}_2=0$) on the $b_3=0$ sub-plane, as demonstrated for PtSr$_5$ in this work.
    }
    \label{fig:topo_vs_normal_DS}
\end{figure}

Dirac semimetals can be categorized into two classes: nonsymmorphic Dirac semimetals  and topological Dirac semimetals \cite{Yang14p4898}. The former class \cite{Young12p140405, Steinberg14p036403, Young15p126803, Wieder16p186402, Wieder18p246}, which arises in PT-symmetric systems, defines a topological critical phase boundary \cite{Murakami07p356} where Dirac points are protected by nonsymmorphic space group symmetries and reside at high-symmetry points at the Brillouin zone (BZ) boundary. Breaking these symmetries allows the system to transition into various topological phases \cite{Wieder16p186402, Bradlyn16paaf5037}, such as nodal-line semimetals \cite{Oh19p201110}, higher-order topological insulators \cite{Benalcazar17p61, Schindler18p918}, or Weyl semimetals \cite{Wan11p205101}. On the other hand, topological Dirac semimetals \cite{Wang12p195320, Wang13p125427, Liu14p677, Wang23p013079} originate from rotational symmetry and involve distinct irreducible representations along a symmetry axis. In this case, band inversion at high-symmetry TRIM points induces Dirac points, but rotational symmetry prevents a full gap opening.

\add{Topological Dirac semimetals are known to host topologically nontrivial insulating phases within certain two-dimensional (2D) momentum subspaces~\cite{Bradlyn17p298, Zhang19p475}.
In contrast, systems that remain topologically trivial across all such subspaces, referred to as normal Dirac semimetals ~\cite{Yang14p4898}, are rarely realized.
Figs.~\ref{fig:topo_vs_normal_DS}(a) and \ref{fig:topo_vs_normal_DS}(b) illustrate two Dirac semimetal scenarios distinguished by the  $\mathbb{Z}_{2}$ index on the time-reversal-invariant $b_{3} = 0$ plane.  
In the conventional case [Fig.~\ref{fig:topo_vs_normal_DS}(a)], a band inversion at a time-reversal-invariant momentum (TRIM) can generate a pair of Dirac points and induces a nontrivial $\mathbb{Z}_{2}$ index on that plane.  
By contrast, in \ce{PtSr5} [Fig.~\ref{fig:topo_vs_normal_DS}(b)], two successive band inversions occur on the same side of the $b_{3}$ axis, producing Dirac points without altering the symmetry eigenvalues at any other TRIM.  
The resulting phase remains topologically trivial, with $\mathbb{Z}_{2}=0$ on all TRIM-containing planes and exhibits no topological crystalline insulating character, thereby constituting a normal Dirac semimetal.
}

\add{The Pt--Sr binary system offers a fertile ground for discovering intermetallic phases stabilized by a delicate interplay of multivalence, strong spin--orbit coupling (SOC), and metallic bonding.   Early reports on Pt-rich compounds such as \ce{Pt5Sr}, \ce{Pt3Sr}, and \ce{Pt2Sr} from the 1950s--60s~\cite{Heumann1957, Wood1958, Bronger1962} motivated exploration of a broader range of stoichiometries toward the Sr-rich regime, including \ce{Pt4Sr5}, \ce{Pt2Sr3}, \ce{Pt3Sr7}, and \ce{PtSr4}~\cite{Palenzona1981}. The compound \ce{PtSr5} itself appears as a distinct entry in the Sr--Pt binary phase diagram~\cite{Palenzona1981}, underscoring its thermodynamic stability as an intermetallic phase.  
This diagram has recently been completed by the AI-guided identification of \ce{PtSr5} as a body-centered tetragonal phase with space group $I4/m$~\cite{Merchant23p80}. Given the strong SOC of Pt, which supports symmetry-protected band crossings, several family members, including \ce{Pt3Sr7}, \ce{Pt4Sr5}, \ce{Pt2Sr}, and \ce{Pt5Sr}, have been classified as topological semimetals in recent catalogues~\cite{Vergniory2019}. The discovery of \ce{PtSr5} therefore naturally raises the question of its topological character.}

In this paper we present first-principles calculations that identify \ce{PtSr5} as the normal Dirac semimetal. In particular, we suggest that \ce{PtSr5} realizes a topologically trivial Dirac semimetal phase, hosting an even number of Dirac points along the $\Gamma$-$Z$ high-symmetry line. Despite the trivial $\mathbb{Z}_2$ topology, \ce{PtSr5} defines the phase boundary of a Weyl semimetal, which we demonstrate by applying an external Zeeman magnetic field along the $z$-axis. Under this perturbation, each Dirac point splits into a pair of Weyl points with opposite Chern numbers, confirming the tunability of its topological phase. This transition is further supported by surface state calculations, revealing the emergence of Weyl cones and modifications in surface states in response to the applied field. Our findings not only establish \ce{PtSr5} as an intriguing platform for studying tunable topological phase transitions but also underscore the effectiveness of AI-driven material discovery in identifying novel quantum materials with unconventional electronic properties.

\section*{Methods}

First-principles density functional theory (DFT) calculations were performed using \textsc{Quantum ESPRESSO} \cite{Giannozzi2009} within the Perdew-Burke-Ernzerhof (PBE) generalized gradient approximation (GGA) \cite{Perdew96p3865} for the exchange-correlation functional. Atomic potentials were modeled using norm-conserving optimized 
pseudopotentials \cite{hamann_optimized_2013} from the {\sc Pseudo-Dojo} library 
\cite{van_setten_pseudodojo_2018}.  Both fully relativistic SOC and non-SOC calculations were conducted to provide a comprehensive analysis of SOC effects on the electronic structure. The plane-wave energy cutoff was set to 350~eV, and self-consistent field calculations were carried out on the Monkhorst-Pack grid of a $7 \times 7 \times 7$ $\mathbf{k}$-mesh \cite{monkhorst_special_1976}, achieving total energy convergence below $10^{-10}$~eV. 

The atomic structure was adopted from the original materials discovery in Ref. \cite{Merchant23p80}, with lattice constants $a_1 = a_2 = a_3 = 7.51$\,\AA. Band structures under an external magnetic field were obtained by first constructing maximally localized Wannier functions with \textsc{Wannier90} \cite{Mostofi14p2309}, employing the density matrix method for disentanglement \cite{Damle15p1463, Damle18p1392}. The wannierization energy window was set to the range $[-2.7, 6.0]$~eV relative to the Fermi level $E = 0$. An external magnetic field $B$ along the $z$-direction was then applied to the SOC bands, treated as an effective Zeeman field that splits the spin-up and spin-down components within the Wannier Hamiltonian. 

Topological properties of \ce{PtSr5} were analyzed using the Fu--Kane parity criterion for $\mathbb{Z}_2$ invariants~\cite{Fu07p045302}, with parity eigenvalues evaluated via \textsc{Qeirrep}~\cite{qeirrep} at the time-reversal-invariant momenta. 
Additional analyses of Dirac and Weyl points, including the computation of Chern numbers and surface Fermi arcs, were carried out using \textsc{WannierTools}~\cite{Wu18p405}. 
The mirror Chern number~\cite{Teo2008surface} was evaluated using Wannier charge center flows~\cite{Rauch21p195103}, as implemented in \textsc{WannierTools}. 
\add{Furthermore, the surface-to-bulk spectral weight ratio was obtained from slab calculations based on the constructed Wannier Hamiltonian incorporating the Zeeman term. 
A slab geometry comprising ten unit cells stacked along the $a_{3}$ direction was modeled in \textsc{WannierTools}, 
and the surface contribution was evaluated by projecting the states onto the top two atomic layers.
}

\section*{Results}

\subsection*{Crystal structure and symmetries of \ce{PtSr5}}

\begin{figure*}[t]
\centering
\includegraphics[width=0.8\linewidth]{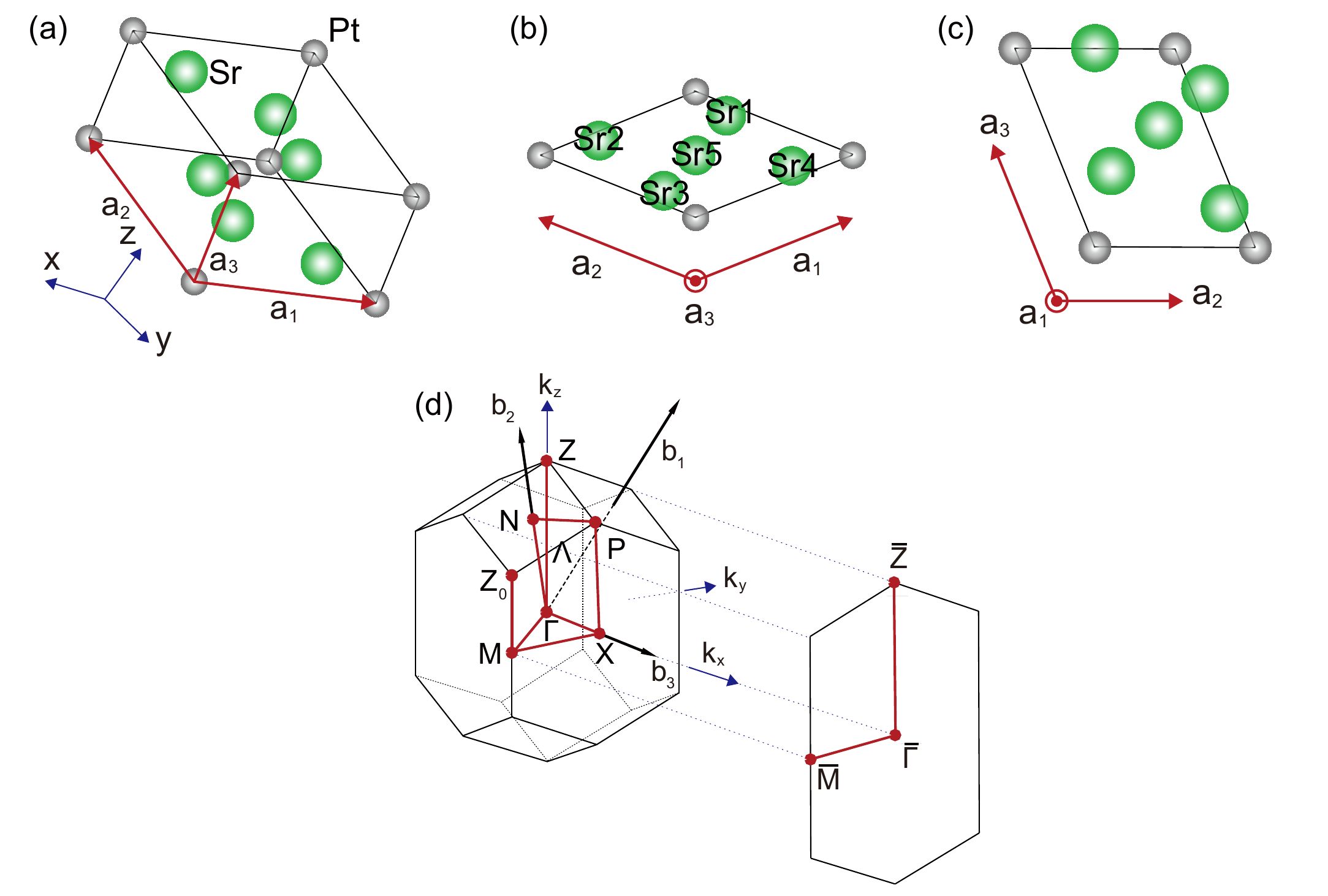}
\caption{\textbf{Crystal structure and Brillouin zone (BZ) of \ce{PtSr5}.}
(a)–(c) Primitive unit cell and lateral views, with red and blue arrows indicating the primitive cell vectors and Cartesian axes, respectively. Pt and Sr atoms are shown as distinct spheres. (d) First BZ, including high-symmetry points (red) and a projected two-dimensional (2D) BZ for the [001] surface orientation.
}
\label{fig:JCm}
\end{figure*}

\ce{PtSr5} crystallizes in a body-centered tetragonal structure. Figs.\,\ref{fig:JCm}(a) and \ref{fig:JCm}(c) present the atomic structure of \ce{PtSr5} with primitive vectors indicated by red arrows. The corresponding primitive unit cell, outlined by the solid box, has lattice constants satisfying $a_1 = a_2 = a_3$ and contains one formula unit comprising five Sr atoms and one Pt atom. Of the five Sr atoms, four [Sr1, Sr2, Sr3, and Sr4 in Fig.\,\ref{fig:JCm}(b)] occupy the Wyckoff position $h$, which is related by the symmetry operations of the $I4/m$ space group. The remaining Sr atom resides at the $b$ position (Sr5), preserving the symmetry individually.

\ce{PtSr5} retains the full symmetry group $I4/m$ (No.\,87), which consists of eight symmetry operations generated by inversion ($\mathcal{P}$), mirror ($m_{a_3}$), four-fold rotation ($C_{4z}$), and a body-centered translation by $(a_1 + a_2 + a_3)/2$. In the conventional tetragonal setting, where $c < a$, the first Brillouin zone (BZ) acquires a hexarhombic dodecahedral shape, as depicted in Fig.\,\ref{fig:JCm}(d).  Important high-symmetry $k$ points are shown in red. Notably, the $\Gamma$--$Z$ line ($\Lambda$, parameterized by $(u,u,-u)$ with free parameter $u$) and the $M$--$Z_0$ line both preserve four-fold rotational symmetry, playing a central role in hosting four-fold-degenerate Dirac points. These points arise through the interplay between rotational symmetry and inversion symmetry and will be discussed in detail later.

For the analysis of topological surface states, we focus on the [001] surface, corresponding to the $k_x$-oriented surface. The projected 2D BZ for this orientation is also illustrated and includes the high-symmetry $\Gamma$--$Z$ and $M$--$Z_0$ lines, which appear horizontally and without oblique distortions. This alignment facilitates a clearer interpretation of the surface states and will be emphasized in subsequent sections.

The previous study indicates that the PtSr$_5$ structure is thermodynamically stable within the resolution of the GNoME calculations~\cite{Merchant23p80}. The reported formation energy per atom is $-0.383$~eV, and the decomposition energy per atom relative to competing phases is $-0.0318$~eV. The negative decomposition energy indicates stability against phase separation. The final GNoME models have a reported mean absolute error of approximately 11~meV/atom for formation energies and a stable prediction accuracy exceeding 80\% when structural information is included~\cite{Merchant23p80}. The combination of negative decomposition energy and fully converged relaxation reported in the dataset is consistent with the stability criteria established in their work. 

\subsection*{Electronic band structure and Dirac points}

\begin{figure}[t]
\centering
\includegraphics[width=1.0\linewidth]{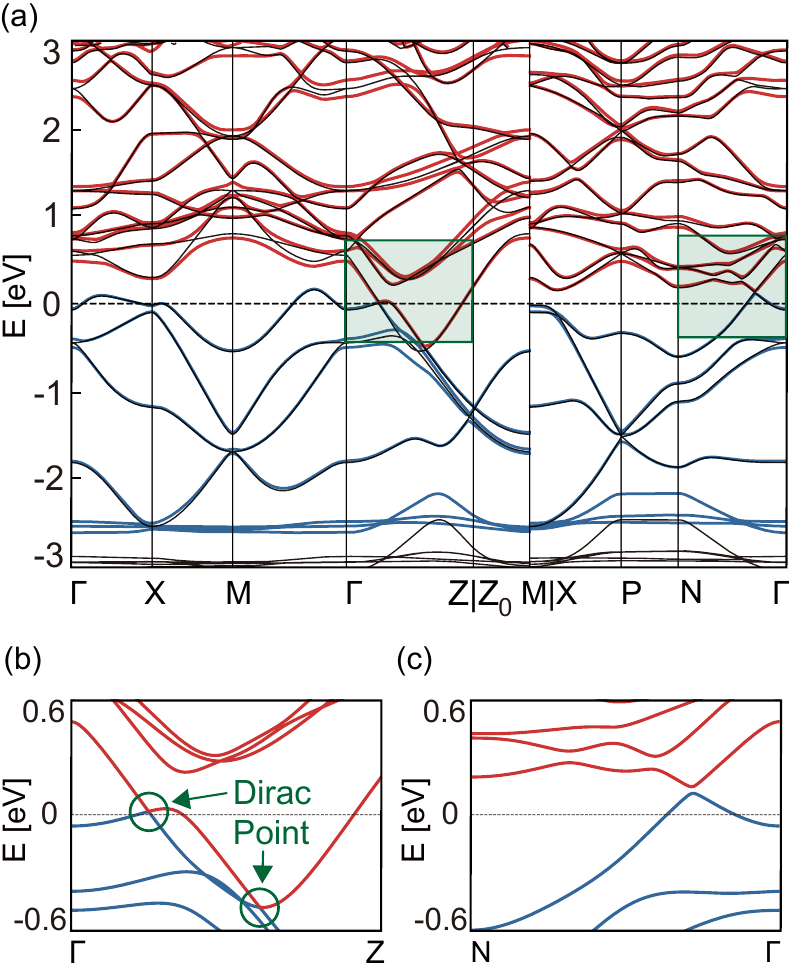}
\caption{\textbf{Electronic energy bands of \ce{PtSr5}.}
(a) Band structures of \ce{PtSr5} with and without spin-orbit coupling (SOC) are shown in color and black, respectively. Blue and red indicate the occupied and unoccupied SOC bands.
(b) Magnified view of two Dirac points (green circles) along the $\Gamma$--$Z$ line, as highlighted by the green box in (a). (c) Magnified view of the small anti-crossing gap (35.8\,meV) along the $N$--$\Gamma$ path, illustrating the SOC-induced band splitting.
}
\label{fig:effham}
\end{figure}

\begin{figure}[t!]
\centering
\includegraphics[width=1.0\linewidth]{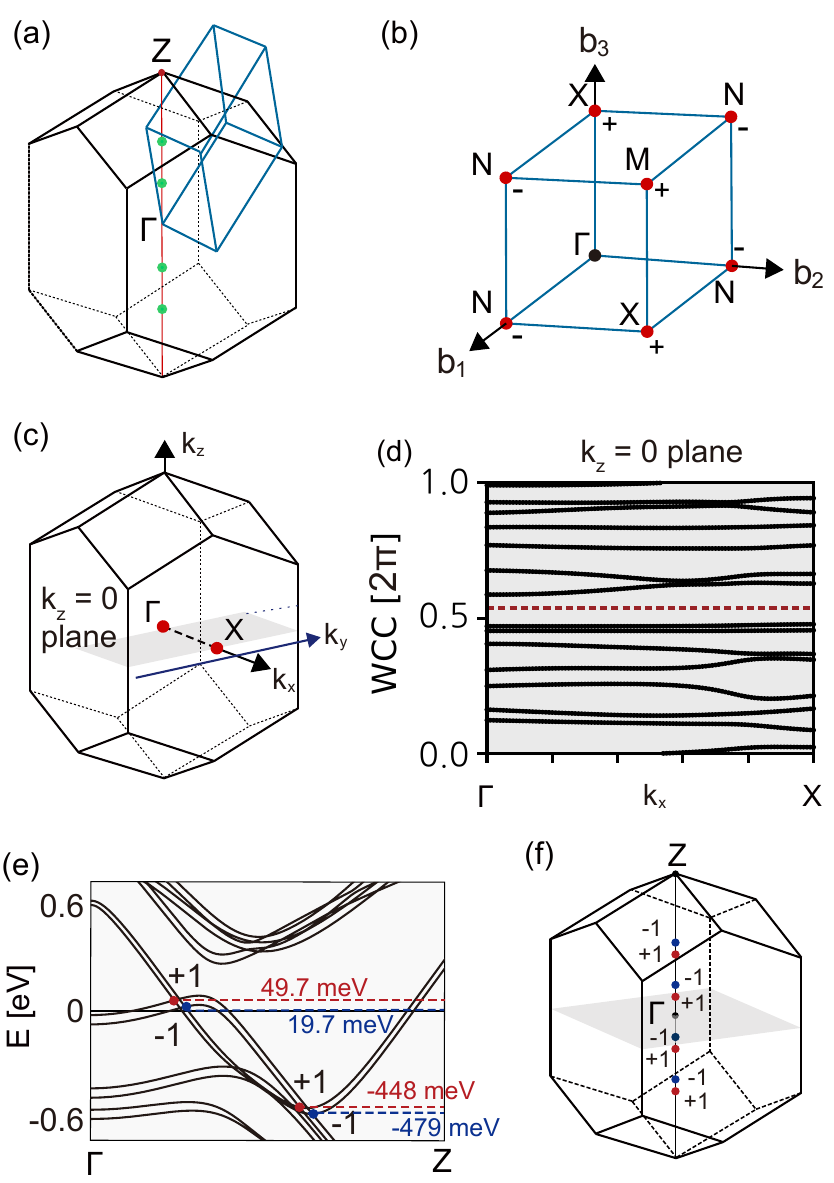}
\caption{\textbf{Dirac points and Weyl transitions in \ce{PtSr5}.}
(a) Schematic illustration of the Dirac points (green dots) and possible Weyl points in momentum space. 
The black shape represents the first BZ, and the red line highlights the high-symmetry $\Gamma$--$Z$ axis. 
A blue cube indicates one-eighth of the BZ (comprising four TRIMs). 
(b) Parity eigenvalue signs ($+$ or $-$) for the eight TRIMs at the corners of a conceptual cube. 
\add{
(c) The $k_z = 0$ plane used for the Wannier charge center (WCC) analysis, where the blue-shaded face denotes the $k_x \ge 0$ half-plane 
and the blue arrow indicates the string direction ($k_y$). 
(d) Evolution of the WCCs obtained by shifting the $k_y$ string along $k_x$ from $\Gamma$ to $X$, 
revealing a well-defined spectral gap (red dashed line) without any winding of charge centers, 
demonstrating that both the $\mathbb{Z}_2$ topological invariant and the mirror Chern number on this plane are trivial.
}
(e) Band structure along $\Gamma$--$Z$ under a $B = 50$\,T field applied along $z$, 
showing each Dirac point splitting into two Weyl points with Chern numbers $+1$ (red) and $-1$ (blue). 
(f) Locations of these Weyl points in the first BZ, with chiralities indicated by red and blue dots.
}
\label{fig:topo}
\end{figure}

\begin{figure*}[t!]
    \centering
    \includegraphics[width=1.0\linewidth]{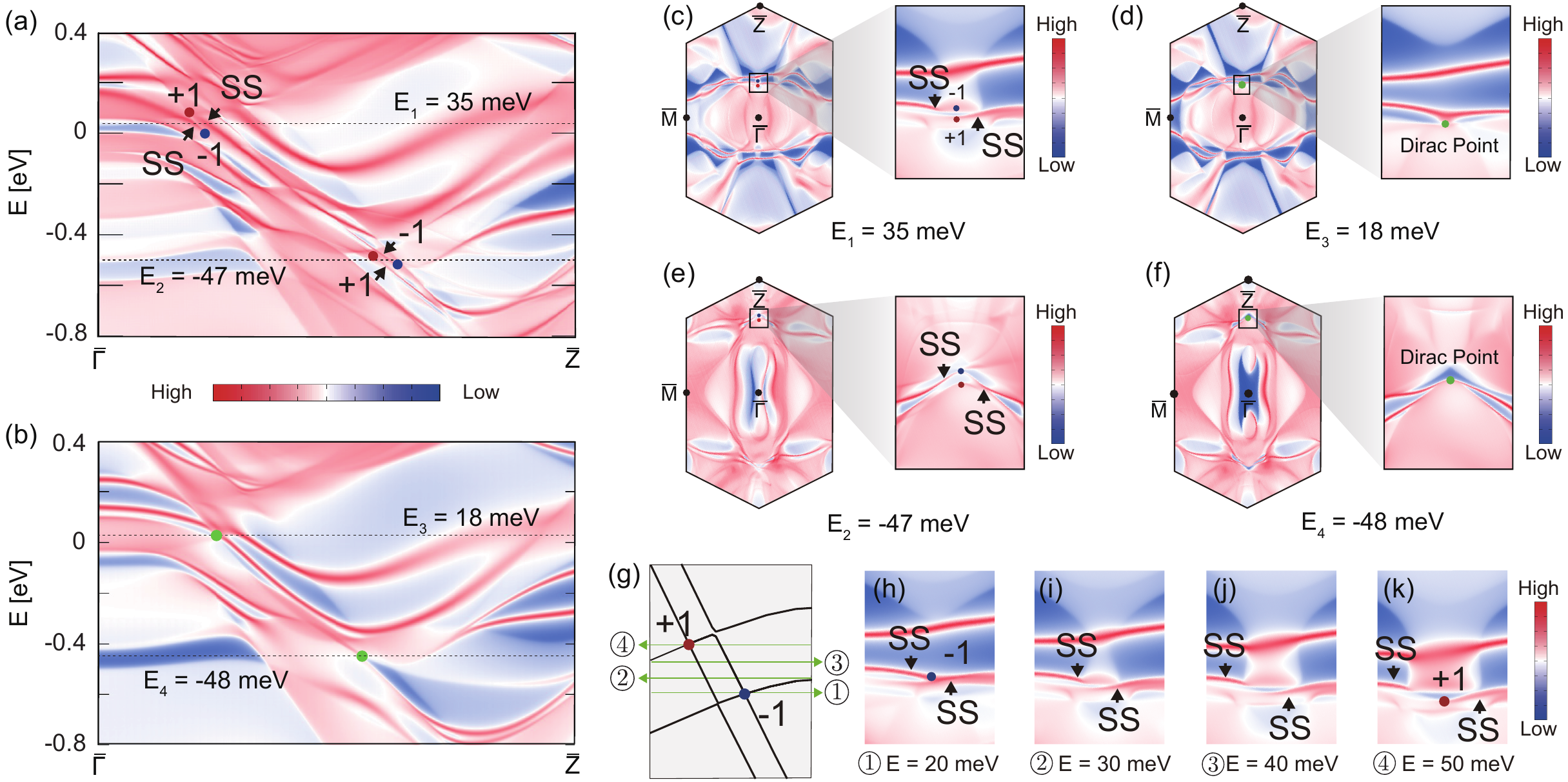}
    \caption{
    \add{
        \textbf{Surface energy spectra revealing the Dirac-to-Weyl transition.}
        (a) and (b) Surface band dispersions along the $\overline{\Gamma}$--$\overline{Z}$ line,
        calculated with (a) and without (b) an external magnetic field $B = 50$\,T.
        Red (blue) circles indicate the Weyl points with Chern numbers $+1$ ($-1$), 
        while green circles denote the Dirac points in the zero-field case.
        The color scale represents the spectral weight projected onto the surface, 
        with red corresponding to stronger surface localization.
        Black arrows in (a) mark the surface states (SS) responsible for the surface Fermi arcs shown in (c) and (e).
        (c) Surface spectral map in the projected 2D BZ at the average energy $E_1 = 35$\,meV
        of the Weyl points in (a), where the momenta of the $+1$ and $-1$ Weyl points are highlighted by red and blue dots.
        (d) Surface spectrum at the Dirac point energy $E_3 = 18$\,meV for the zero-field case,
        with the magnified inset highlighting the Dirac point (green circle).
        (e) and (f) show surface spectra at $E_2 = -47$\,meV and $E_4 = -48$\,meV, respectively, under $B = 50$\,T.
        (g) Schematic energy–momentum diagram illustrating the four constant-energy planes corresponding to panels (h)–(k), where (h) $E=20$\,meV, (i) $E=30$\,meV, (j) $E=40$\,meV, and (k) $E=50$\,meV.
        The progression from (h) to (k) visualizes the continuous Fermi arc connecting the $-1$ Weyl node to the $+1$ Weyl node.
        The black arrow traces the topological SS Fermi arc.
    }
    }
    \label{fig:Dirac_to_Weyl}
\end{figure*}

Figure\,\ref{fig:effham} presents the electronic energy band structure of \ce{PtSr5}, obtained from DFT calculations. The system exhibits a compensated semimetal phase, wherein conduction and valence bands overlap the Fermi level. Owing to the combined inversion $(\mathcal{P})$ and time-reversal $(\mathcal{T})$ symmetry such that $(\mathcal{PT})^2 = -1$, all bands form Kramers-degenerate pairs. Along the $\Gamma$--$Z$ high-symmetry line, two doubly degenerate bands cross, generating fourfold degenerate Dirac points, as highlighted in Fig.\,\ref{fig:effham}(b). These Dirac points reside at $+18\,\mathrm{meV}$ and $-484\,\mathrm{meV}$ relative to the Fermi level. In contrast, all other $\mathbf{k}$-points in the Brillouin zone (BZ) show a direct gap between the $N$ and $N+1$ bands. For instance, along the $N$--$\Gamma$ path, the bands appear to cross but form a small anti-crossing with a direct gap of $35.8\,\mathrm{meV}$, as indicated in Fig.\,\ref{fig:effham}(c). These Dirac points are thus expected to influence the low-energy physics of \ce{PtSr5}.

Unlike typical topological Dirac semimetals, \ce{PtSr5} hosts a pair of Dirac points in half the BZ, leading to four Dirac points in the first BZ. As illustrated in Fig.\,\ref{fig:effham}(b), two appear in the upper half of the bulk BZ; by inversion symmetry, the other two lie along the opposite $\Gamma$--$Z$ line, totaling four Dirac points. In many topological Dirac semimetals, an odd number of Dirac points per half BZ often stems from band inversion at TRIMs, leading to a nontrivial topology. Here, however, the Dirac crossings occur in the interior of the BZ, suggesting a trivial phase in each 2D subspace. 

Moreover, a single Dirac crossing along a high-symmetry line that intersects distinct TRIM planes (each containing four TRIMs, such as $\Gamma$, $X$, and $M$ points) can signal band inversion at a single TRIM, commonly associated with a topological Dirac semimetal. In that scenario, one might observe a Dirac point solely in the $k_z>0$ or $k_z<0$ sector, reflecting a topologically nontrivial phase. In contrast, \ce{PtSr5} exhibits two such Dirac crossings per half BZ and thus yields four points in total. This configuration indicates that no single TRIM plane undergoes band inversion alone, implying that \ce{PtSr5} realizes a normal Dirac semimetal phase without hosting a local topological insulating index. We verify this point more rigorously in the following section.

\add{
It is also instructive to examine the effects of spin–orbit coupling (SOC) on the band structure. As shown in Fig.~\ref{fig:effham}, the bands calculated without SOC are plotted as black lines. A close inspection reveals that, in the absence of SOC, the only band crossings between the $N$ and $N+1$ bands occur along the $\Gamma$–$Z$ axis, with one pair on each side of the BZ. Similar to the SOC-on case, these crossings are protected by rotational symmetry along $\Gamma$–$Z$, as the two bands involved belong to distinct symmetry representations. Including SOC deepens the band inversion, and the sizable magnitude of SOC in \ce{PtSr5} is evident from the clear splitting between the SOC-on and SOC-off band structures. 
These observations clarify that SOC primarily enhances the existing band inversion without altering the overall band topology, consistent with the normal Dirac semimetal nature discussed below.
}

\subsection*{Trivial band topology}

To confirm that \ce{PtSr5} realizes locally trivial topology within any 2D time-reversal-invariant plane, we compute the $\mathbb{Z}_2$ invariants for all such planes in the three-dimensional (3D) Brillouin zone (BZ). As illustrated in Fig.\,\ref{fig:topo}(a), the Dirac points (DPs) lie outside these time-reversal invariant planes, ensuring well-defined weak indices. This is in sharp contrast to conventional topological Dirac semimetals, whose DPs often reside within a time-reversal invariant plane, making the $\mathbb{Z}_2$ indices ill-defined. In \ce{PtSr5}, none of the high-symmetry fourfold rotational axes hosting Dirac points intersect any 2D subspace containing four TRIMs, so we can unambiguously calculate the weak indices.

We find that every time-reversal invariant plane exhibits a trivial $\mathbb{Z}_2$ invariant of $+1$. Concretely, we examine the eight TRIMs [Fig.\,\ref{fig:topo}(b)] and compute the product of parity eigenvalues for the occupied states in each plane containing four of these TRIM points. Each such product is $+1$, leading to vanishing weak indices $(v_1, v_2, v_3) = (0,0,0)$. Consequently, \ce{PtSr5} supports Dirac points without developing a nontrivial band topology, earning the designation of a trivial topological Dirac semimetal. 
\add{We further confirm this result by examining the mirror-invariant $k_z = 0$ plane shown in Fig.\,\ref{fig:topo}(c), where the mirror Chern number~\cite{Teo2008surface, Rauch21p195103} is evaluated from the Wannier charge center (WCC) evolution in Fig.\,\ref{fig:topo}(d). The WCC spectrum exhibits a well-defined global gap (highlighted by the red dashed line) without any spectral flow across the Brillouin zone, demonstrating that both the $\mathbb{Z}_2$ invariant and the mirror Chern number are trivial on this plane. These results collectively reinforce the normal Dirac semimetal nature of \ce{PtSr5}.}
This even number of Dirac points per half BZ, arising away from the 2D time-reversal invariant planes, reinforces the conclusion that no topological index emerges from these band crossings. 

Given that each Dirac cone can be viewed as a pair of Weyl cones with opposite chirality, we next explore whether breaking time-reversal symmetry can split these Dirac points into Weyl points. The two Dirac points along the $\Gamma$--$Z$ axis suggest four Weyl points could appear if $\mathcal{T}$ symmetry is lifted, a possibility we examine in the following section.

\begin{figure}[t]
    \centering
    \includegraphics[width=1.0\linewidth]{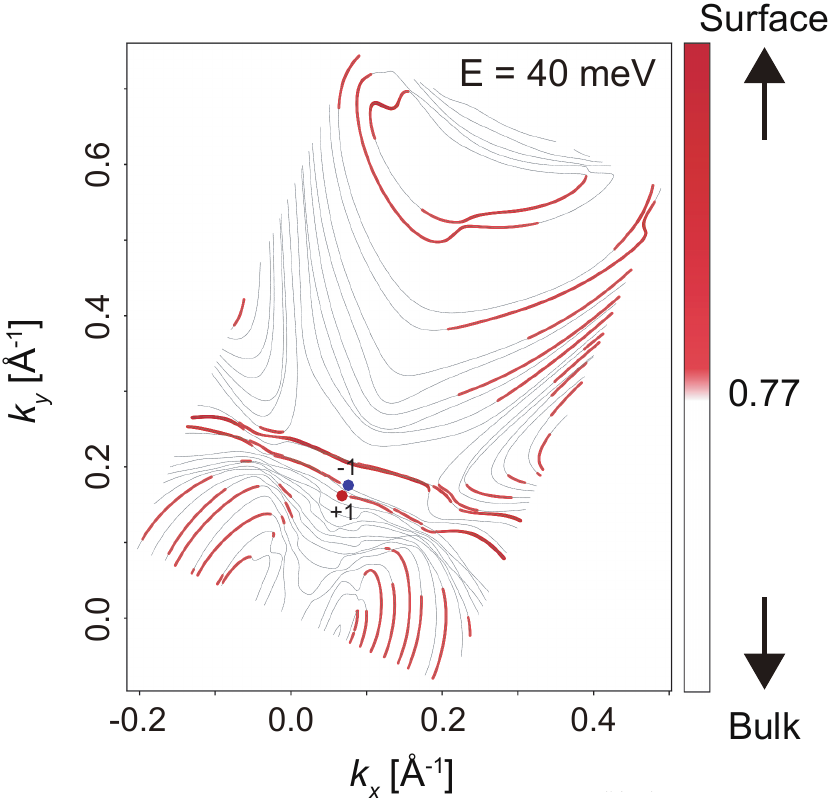}
    \caption{\add{\textbf{Surface Projection of Zeeman-Induced Weyl Arcs.}
        Constant-energy contour of the slab electronic structure of PtSr$_5$ at $E = E_F + 40~\mathrm{meV}$,
        obtained from a 10-unit-cell slab calculation with vacuum spacing.
        The color scale indicates the bulk-to-surface spectral weight ratio, 
        with red regions denoting surface-dominant states ($>$77\% surface contribution).
        The red and blue circles denote Weyl points with Chern numbers $+1$ and $-1$, 
        located near $E = 50$ and $20~\mathrm{meV}$, respectively, and connected by a surface Fermi arc at this intermediate energy.
    }
    }
    \label{fig:surface_ratio}
\end{figure}

\subsection*{Field-induced Weyl points and Chern numbers}

\add{While the topological phase of \ce{PtSr5} appears trivial, the presence of Dirac points in momentum space defines a topological phase boundary that allows nontrivial topological phenomena to occur when a phase transition is induced.} When TRS is explicitly broken, a pair of Weyl points with opposite Chern numbers can emerge from each Dirac point. To demonstrate this, we apply a Zeeman magnetic field along the $z$-axis (approximately 50\,T, with a Land\'e $g$-factor of 15) and calculate both the bulk band structure and surface states of \ce{PtSr5}.%
\add{We emphasize that the relatively large $B$ field and $g$ factor used here are chosen solely for numerical clarity, without loss of generality. Importantly, the same Dirac-to-Weyl node splitting mechanism has been experimentally realized at much lower magnetic fields—for instance, in Bi$_{0.96}$Sb$_{0.04}$ under a 9,T field, where clear transport signatures of the chiral anomaly were observed~\cite{Shin2017}. Moreover, giant effective $g$ factors exceeding $10^2$--$10^4$ have been achieved in engineered InAsSb/InSb superlattices~\cite{Jiang2022}, validating the physical plausibility of the Zeeman coupling strength assumed in our calculations.}

Fig.\,\ref{fig:topo}(e) shows the band structure under the field. The doubly degenerate Kramer pairs split into two branches, between which can form band crossing. This result shows that eight gapless points arise from the Zeeman splitting of the bands. Four of these points lie along the $\Gamma$--$Z$ line, and the other four appear symmetrically along the opposite line, as illustrated in Fig.\,\ref{fig:topo}(f). They originate from the splitting of the Dirac points, with the Weyl point pairs located at 49.7\,meV and 19.7\,meV above the Fermi level, and at -448\,meV and -479\,meV below.

We confirm their Weyl nature by computing Chern numbers. Among the eight points, four have Chern numbers of $+1$, and four have $-1$. Each Weyl pair, carrying opposite Chern numbers, emerges from a single Dirac point. Notably, these Weyl pairs remain close to the original Dirac points and align along the applied field, indicating that the Zeeman effect lifts the Dirac degeneracy and produces pairs of Weyl cones with opposite chirality. Hence, \ce{PtSr5} can be driven from a Dirac semimetal phase into a Weyl semimetal by an external magnetic field, illustrating the tunability of its topological properties.

\subsection*{Field-induced topological Fermi arcs}

Having established the emergence of Weyl points under a magnetic field, we next examine the corresponding surface states to identify any Fermi arcs indicative of a Weyl semimetal phase. Figs.\,\add{\ref{fig:Dirac_to_Weyl}}(a) and (b) show the surface band dispersions along the $\overline{\Gamma}$--$\overline{Z}$ line, computed with and without a $B=50$\,T Zeeman field, respectively. Under the external $B$-field [Fig.\,\add{\ref{fig:Dirac_to_Weyl}}(a)], the overall surface spectral weight becomes more dispersed in both energy and momentum compared to the zero-field case [Fig.\,\add{\ref{fig:Dirac_to_Weyl}}(b)]. This broadened appearance reflects the Zeeman splitting of bulk states, as seen in the bulk band structure of Fig.\,\add{\ref{fig:topo}(c)}, in which the Dirac point along $\Gamma$--$Z$ is replaced by two Weyl points at different energies, indicating that spin-up and spin-down bands are no longer degenerate. Consequently, the surface energy spectra inherit this splitting: instead of one narrow Dirac-like intersection, two broader branches emerge at slightly different energies and momenta. These changes are most evident in the high-intensity regions that spread farther apart in energy relative to the Fermi level, mirroring the bulk spin polarization induced by the Zeeman field.

In addition to the bulk states, their magnified views reveal the surface localized states, including topological Fermi arcs. Figs.\,\add{\ref{fig:Dirac_to_Weyl}(c)–(f)} present the projected 2D BZ at fixed energies, allowing us to see the locations of these Weyl and Dirac points more clearly. In Fig.\,\add{\ref{fig:Dirac_to_Weyl}(c)}, taken at $E=\add{E_1=35}$\,meV (the average energy of the two Weyl points), red and blue dots mark the Weyl nodes with Chern numbers $+1$ and $-1$, respectively. From each Weyl node, a bright surface branch emerges, consistent with a Fermi arc characteristic of the Weyl semimetal phase. \add{At $E=\!E_2=-47$\,meV [Fig.\,\ref{fig:Dirac_to_Weyl}(e)], a lower-energy arc is visible as a seemingly disconnected segment; this segment continuously evolves into the connected arc at $E_1$ as the energy is raised. The four constant-energy cuts in Figs.\,\ref{fig:Dirac_to_Weyl}(g)–(k) ($E=$20,30,40, and 50\,meV) quantify this evolution: starting from the $-1$ Weyl node at 20\,meV (h), the arc progressively hybridizes with the emerging electron pocket (i–j), becoming quasi-resonant and partially delocalized in the bulk, and finally reconnects to the $+1$ Weyl node at 50\,meV (k).} This observation directly confirms that an external field drives \ce{PtSr5} from a trivial Dirac semimetal to a Weyl semimetal. 

In contrast, when no $B$-field is present, the apparent Fermi arc states become connected at the Dirac point and thus lose their topological character. Fig.\,\add{\ref{fig:Dirac_to_Weyl}(d)} shows the surface energy spectrum in the same region of the BZ at zero field, plotted 18\,meV above the Fermi level to highlight the Dirac point (white dot). Two otherwise disconnected branches meet at this point. Although the system is globally topologically trivial, a strongly localized surface state still intersects the Dirac point, indicating that \ce{PtSr5} lies on the boundary of a Weyl semimetal phase and exhibits surface features reminiscent of true Fermi arcs. Because the Dirac semimetal here is not protected by a nontrivial index, one expects an even number of surface branches and no robust topological protection in the absence of $B$-field. \add{Consistently, the zero-field dispersion in Fig.\,\ref{fig:Dirac_to_Weyl}(b) shows an even number of surface branches along $\overline{\Gamma}$--$\overline{Z}$, in agreement with the bulk triviality and the surface-band counting inferred from Fig.\,\ref{fig:topo}(d).} Nevertheless, the presence of such surface states in this trivial Dirac phase highlights the intriguing topological aspect of \ce{PtSr5} as a normal Dirac semimetal that can readily transition to a Weyl semimetal under symmetry breaking.

\add{To quantitatively confirm the surface localization of the arc, we further computed the surface-to-bulk spectral-weight ratio using a ten-unit-cell slab based on the Wannier Hamiltonian including the Zeeman term. The surface contribution was evaluated by projecting the states onto the top two atomic layers. As shown in Fig.\,\ref{fig:surface_ratio}, the arc near $E=40$\,meV exhibits a surface weight of $\sim\!77\%$, corroborating its strongly surface-localized character despite the intermediate hybridization with bulk pockets.}

\section*{Conclusion}

We have investigated the intermetallic compound \ce{PtSr5}, known as an AI-driven material, and demonstrated that it hosts a Dirac semimetal phase with a trivial $\mathbb{Z}_2$ index. First-principles density functional theory calculations, combined with symmetry indicator analysis, confirm that the Dirac points in \ce{PtSr5} do not give rise to a nontrivial topological invariant. Nevertheless, we find that applying a Zeeman magnetic field splits each Dirac point into a pair of Weyl points with opposite chiralities, thus revealing a tunable Weyl semimetal phase. 
\add{Our surface-state analysis demonstrates that the resulting Weyl nodes are connected by Fermi arcs whose energy evolution, from disconnected to reconnected branches, traces the field-driven topological surface states as a function of energy. 
Furthermore, slab calculations show that these arc states possess a surface localization ratio of about 77\%, providing quantitative evidence for their surface origin.} 
This transition, corroborated by Chern number calculations and surface-state analyses, highlights the interplay between symmetry, magnetic fields, and band topology in \ce{PtSr5}.

Our work underscores the effectiveness of data-driven approaches in identifying promising quantum materials, extending the catalog of Dirac and Weyl semimetals to previously uncharted compounds. While \ce{PtSr5} offers a platform to study the evolution of Dirac fermions into Weyl fermions, many other candidate materials discovered through AI-driven strategies may similarly enrich the phase diagram of topological matter. We anticipate that further integration of high-throughput computations, machine learning, and experimental validation will accelerate the discovery of exotic electronic states, paving the way for new fundamental insights and potential technological applications.

\section*{Acknowledgments}

This work was supported by the National Research Foundation of Korea (NRF) grant funded by the Korea government(MSIT) (RS-2025-21772968 and No. RS-2019-NR040081). The Korea Institute of Science and Technology Information (KISTI) (KSC-2025- CRE-0058) provided the computational resource. 
I.L. and C.L. performed the DFT calculations and topological analyses. Y.K. conceived and supervised the project. All authors discussed the results and contributed to writing the manuscript. I.L. and C.L. contributed equally to this work.

\def\urlprefix{}
\def\href#1#2{#2}

\balance
\bibliographystyle{elsarticle-num} 

\begin{thebibliography}{10}
\expandafter\ifx\csname url\endcsname\relax
  \def\url#1{\texttt{#1}}\fi
\expandafter\ifx\csname urlprefix\endcsname\relax\def\urlprefix{URL }\fi
\expandafter\ifx\csname href\endcsname\relax
  \def\href#1#2{#2} \def\path#1{#1}\fi

\bibitem{Merchant23p80}
A.~Merchant, S.~Batzner, S.~S. Schoenholz, M.~Aykol, G.~Cheon, E.~D. Cubuk, \href{http://dx.doi.org/10.1038/s41586-023-06735-9}{Scaling deep learning for materials discovery}, Nature 624~(7990) (2023) 80–85.
\newblock \href {https://doi.org/10.1038/s41586-023-06735-9} {\path{doi:10.1038/s41586-023-06735-9}}.
\newline\urlprefix\url{http://dx.doi.org/10.1038/s41586-023-06735-9}

\bibitem{Kane05p226801}
C.~L. Kane, E.~J. Mele, Quantum spin hall effect in graphene, Physical review letters 95~(22) (2005) 226801.

\bibitem{Bernevig06p1757}
B.~A. Bernevig, T.~L. Hughes, S.-C. Zhang, Quantum spin hall effect and topological phase transition in hgte quantum wells, science 314~(5806) (2006) 1757--1761.

\bibitem{Fu07p045302}
L.~Fu, C.~L. Kane, Topological insulators with inversion symmetry, Physical Review B—Condensed Matter and Materials Physics 76~(4) (2007) 045302.

\bibitem{Hasan10p3045}
M.~Z. Hasan, C.~L. Kane, Colloquium: Topological insulators, Rev. Mod. Phys. 82 (2010) 3045.

\bibitem{Qi11p1057}
X.-L. Qi, S.-C. Zhang, Topological insulators and superconductors, Rev. Mod. Phys. 83 (2011) 1057.

\bibitem{Armitage18p015001}
N.~P. Armitage, E.~J. Mele, A.~Vishwanath, Weyl and dirac semimetals in three-dimensional solids, Rev. Mod. Phys. 90 (2018) 015001.

\bibitem{Xu11p186806}
G.~Xu, H.~Weng, Z.~Wang, X.~Dai, Z.~Fang, Chern semimetal and the quantized anomalous hall effect in ${\mathrm{hgcr}}_{2}{\mathrm{se}}_{4}$, Phys. Rev. Lett. 107 (2011) 186806.

\bibitem{Xu15p613}
S.-Y. Xu, I.~Belopolski, N.~Alidoust, M.~Neupane, G.~Bian, C.~Zhang, R.~Sankar, G.~Chang, Z.~Yuan, C.-C. Lee, S.-M. Huang, H.~Zheng, J.~Ma, D.~S. Sanchez, B.~Wang, A.~Bansil, F.~Chou, P.~P. Shibayev, H.~Lin, S.~Jia, M.~Z. Hasan, Discovery of a weyl fermion semimetal and topological fermi arcs, Science 349 (2015) 613.

\bibitem{Huang15p7373}
S.-M. Huang, S.-Y. Xu, I.~Belopolski, C.-C. Lee, G.~Chang, B.~Wang, N.~Alidoust, G.~Bian, M.~Neupane, C.~Zhang, S.~Jia, A.~Bansil, H.~Lin, M.~Z. Hasan, A weyl fermion semimetal with surface fermi arcs in the transition metal monopnictide taas class, Nat. Commun. 6 (2015) 7373.

\bibitem{Weng15p011029}
H.~Weng, C.~Fang, Z.~Fang, B.~A. Bernevig, X.~Dai, Weyl semimetal phase in noncentrosymmetric transition-metal monophosphides, Phys. Rev. X 5 (2015) 011029.

\bibitem{Gao19p153}
H.~Gao, J.~W. Venderbos, Y.~Kim, A.~M. Rappe, Topological semimetals from first principles, Annual Review of Materials Research 49~(1) (2019) 153--183.

\bibitem{Lv15p031013}
B.~Q. Lv, N.~Xu, H.~M. Weng, J.~Z. Ma, P.~Richard, X.~C. Huang, L.~X. Zhao, G.~F. Chen, C.~E. Matt, F.~Bisti, V.~N. Strocov, J.~Mesot, Z.~Fang, X.~Dai, T.~Qian, M.~Shi, H.~Ding, Observation of weyl nodes in taas, Phys. Rev. X 5 (2015) 031013.

\bibitem{Liu14p864}
Z.~Liu, B.~Zhou, Y.~Zhang, Z.~Wang, H.~Weng, D.~Prabhakaran, S.-K. Mo, Z.~Shen, Z.~Fang, X.~Dai, et~al., Discovery of a three-dimensional topological dirac semimetal, na3bi, Science 343~(6173) (2014) 864--867.

\bibitem{Liu14p677}
Z.~K. Liu, J.~Jiang, B.~Zhou, Z.~Wang, Y.~Zhang, H.~Weng, D.~Prabhakaran, S.-K. Mo, H.~Peng, P.~Dudin, T.~Kim, M.~Hoesch, Z.~Fang, X.~Dai, Z.~Shen, D.~Feng, Z.~Hussain, Y.~L. Chen, A stable three-dimensional topological dirac semimetal cd$_3$as$_2$, Nat. Mater. 13 (2014) 677.

\bibitem{Neupane14p3786}
M.~Neupane, S.-Y. Xu, R.~Sankar, N.~Alidoust, G.~Bian, A.~Neupane, C.~Zhang, K.~Segawa, R.~Sankar, J.~Zhujun, C.~Liu, I.~Belopolski, T.-R. Chang, H.-T. Jeng, H.~Lin, A.~Bansil, F.~Chou, M.~Z. Hasan, Observation of a three-dimensional topological dirac semimetal phase in high-mobility cd$_3$as$_2$, Nat. Commun. 5 (2014) 3786.

\bibitem{Deng16p1105}
K.~Deng, G.~Wan, P.~Deng, K.~Zhang, S.~Ding, E.~Wang, M.~Yan, H.~Huang, H.~Zhang, Z.~Xu, J.~Denlinger, A.~Fedorov, H.~Yang, W.~Duan, H.~Yao, Y.~Wu, Y.~Fan, S.~Zhang, S.~Zhou, Experimental observation of topological fermi arcs in type-ii weyl semimetal mote$_2$, Nat. Phys. 12 (2016) 1105.

\bibitem{Soluyanov15p495}
A.~A. Soluyanov, D.~Gresch, Z.~Wang, Q.~Wu, M.~N. Ali, E.~Talantsev, X.~Zhu, N.~Kumar, B.~Bradlyn, A.~Cairns, K.~Frohna, G.~Osterhoudt, D.~Low, Y.~Shu, D.~S. Sanchez, R.~Sankar, F.~Chou, T.~Neupert, R.~J. Cava, B.~A. Bernevig, Type-ii weyl semimetals, Nature 527 (2015) 495.

\bibitem{Wang12p195320}
Z.~Wang, Y.~Sun, X.-Q. Chen, C.~Franchini, G.~Xu, H.~Weng, X.~Dai, Z.~Fang, Dirac semimetal and topological phase transitions in $a_3$bi ($a$ = na, k, rb), Phys. Rev. B 85 (2012) 195320.

\bibitem{Wang13p125427}
Z.~Wang, H.~Weng, Q.~Wu, X.~Dai, Z.~Fang, Three-dimensional dirac semimetal and quantum transport in ${\mathrm{cd}}_{3}{\mathrm{as}}_{2}$, Phys. Rev. B 88 (2013) 125427.

\bibitem{Vafek14p83}
O.~Vafek, A.~Vishwanath, Dirac fermions in solids: From high-${T}_c$ cuprates and graphene to topological insulators and weyl semimetals, Annu. Rev. Condens. Matter Phys. 5 (2014) 83.

\bibitem{Nielsen83p389}
H.~B. Nielsen, M.~Ninomiya, The adler-bell-jackiw anomaly and weyl fermions in a crystal, Phys. Lett. B 130 (1983) 389.

\bibitem{Son13p104412}
D.~T. Son, B.~Z. Spivak, Chiral anomaly and classical negative magnetoresistance of weyl metals, Phys. Rev. B 88 (2013) 104412.

\bibitem{Xiong15p413}
J.~Xiong, S.~K. Kushwaha, T.~Liang, J.~W. Krizan, M.~Hirschberger, W.~Wang, R.~J. Cava, N.~P. Ong, Evidence for the chiral anomaly in the dirac semimetal na$_3$bi, Science 350 (2015) 413.

\bibitem{Yang14p4898}
B.-J. Yang, N.~Nagaosa, Classification of stable three-dimensional dirac semimetals with nontrivial topology, Nat. Commun. 5 (2014) 4898.

\bibitem{Young12p140405}
S.~M. Young, S.~Zaheer, J.~C.~Y. Teo, C.~L. Kane, E.~J. Mele, A.~M. Rappe, Dirac semimetal in three dimensions, Phys. Rev. Lett. 108 (2012) 140405.

\bibitem{Steinberg14p036403}
J.~A. Steinberg, S.~M. Young, S.~Zaheer, C.~Kane, E.~Mele, A.~M. Rappe, Bulk dirac points in distorted spinels, Physical review letters 112~(3) (2014) 036403.

\bibitem{Young15p126803}
S.~M. Young, C.~L. Kane, \href{https://link.aps.org/doi/10.1103/PhysRevLett.115.126803}{Dirac semimetals in two dimensions}, Phys. Rev. Lett. 115 (2015) 126803.
\newblock \href {https://doi.org/10.1103/PhysRevLett.115.126803} {\path{doi:10.1103/PhysRevLett.115.126803}}.
\newline\urlprefix\url{https://link.aps.org/doi/10.1103/PhysRevLett.115.126803}

\bibitem{Wieder16p186402}
B.~J. Wieder, Y.~Kim, A.~Rappe, C.~Kane, Double dirac semimetals in three dimensions, Physical review letters 116~(18) (2016) 186402.

\bibitem{Wieder18p246}
B.~J. Wieder, B.~Bradlyn, Z.~Wang, J.~Cano, Y.~Kim, H.-S.~D. Kim, A.~M. Rappe, C.~Kane, B.~A. Bernevig, Wallpaper fermions and the nonsymmorphic dirac insulator, Science 361~(6399) (2018) 246--251.

\bibitem{Murakami07p356}
S.~Murakami, Phase transition between the quantum spin hall and insulator phases in 3d: emergence of a topological gapless phase, New J. Phys. 9 (2007) 356.

\bibitem{Bradlyn16paaf5037}
B.~Bradlyn, J.~Cano, Z.~Wang, M.~G. Vergniory, C.~Felser, R.~J. Cava, B.~A. Bernevig, Beyond dirac and weyl fermions: Unconventional quasiparticles in conventional crystals, Science 353 (2016) aaf5037.

\bibitem{Oh19p201110}
Y.-T. Oh, H.-G. Min, Y.~Kim, \href{https://link.aps.org/doi/10.1103/PhysRevB.99.201110}{Dual topological nodal line and nonsymmorphic dirac semimetal in three dimensions}, Phys. Rev. B 99 (2019) 201110.
\newblock \href {https://doi.org/10.1103/PhysRevB.99.201110} {\path{doi:10.1103/PhysRevB.99.201110}}.
\newline\urlprefix\url{https://link.aps.org/doi/10.1103/PhysRevB.99.201110}

\bibitem{Benalcazar17p61}
W.~A. Benalcazar, B.~A. Bernevig, T.~L. Hughes, Quantized electric multipole insulators, Science 357 (2017) 61.

\bibitem{Schindler18p918}
F.~Schindler, Z.~Wang, M.~G. Vergniory, A.~M. Cook, A.~Murani, S.~Sengupta, A.~Kasumov, R.~Deblock, S.~Jeon, I.~Drozdov, H.~Bouchiat, S.~Gu\'eron, A.~Yazdani, B.~A. Bernevig, T.~Neupert, Higher-order topology in bismuth, Nat. Phys. 14 (2018) 918.

\bibitem{Wan11p205101}
X.~Wan, A.~M. Turner, A.~Vishwanath, S.~Y. Savrasov, Topological semimetal and fermi-arc surface states in the electronic structure of pyrochlore iridates, Phys. Rev. B 83 (2011) 205101.

\bibitem{Wang23p013079}
Q.~Wang, M.~Abeykoon, D.~Graf, Y.~Liu, H.~Lei, J.~Ma, M.~Shi, B.~Yan, C.~Petrovic, Topological dirac semimetal baausb, Phys. Rev. Research 5 (2023) 013079.

\bibitem{Bradlyn17p298}
B.~Bradlyn, L.~Elcoro, J.~Cano, M.~G. Vergniory, Z.~Wang, C.~Felser, M.~I. Aroyo, B.~A. Bernevig, Topological quantum chemistry, Nature 547 (2017) 298.

\bibitem{Zhang19p475}
T.~Zhang, Y.~Jiang, Z.~Song, H.~Huang, Y.~He, Z.~Fang, H.~Weng, C.~Fang, Catalogue of topological electronic materials, Nature 566 (2019) 475.

\bibitem{Heumann1957}
T.~Heumann, M.~Kniepmeyer, \href{http://dx.doi.org/10.1002/zaac.19572900309}{A5b‐phasen vom typ cu5ca und lavesphasen in den systemen des strontiums mit palladium, platin, rhodium und iridium}, Zeitschrift f\"{u}r anorganische und allgemeine Chemie 290~(3–4) (1957) 191–204.
\newblock \href {https://doi.org/10.1002/zaac.19572900309} {\path{doi:10.1002/zaac.19572900309}}.
\newline\urlprefix\url{http://dx.doi.org/10.1002/zaac.19572900309}

\bibitem{Wood1958}
E.~A. Wood, V.~B. Compton, \href{http://dx.doi.org/10.1107/S0365110X58001134}{Laves-phase compounds of alkaline earths and noble metals}, Acta Crystallographica 11~(6) (1958) 429–433.
\newblock \href {https://doi.org/10.1107/s0365110x58001134} {\path{doi:10.1107/s0365110x58001134}}.
\newline\urlprefix\url{http://dx.doi.org/10.1107/S0365110X58001134}

\bibitem{Bronger1962}
W.~Bronger, W.~Klemm, \href{http://dx.doi.org/10.1002/zaac.19623190110}{Darstellung von legierungen des platins mit unedlen metallen}, Zeitschrift f\"{u}r anorganische und allgemeine Chemie 319~(1–2) (1962) 58–81.
\newblock \href {https://doi.org/10.1002/zaac.19623190110} {\path{doi:10.1002/zaac.19623190110}}.
\newline\urlprefix\url{http://dx.doi.org/10.1002/zaac.19623190110}

\bibitem{Palenzona1981}
A.~Palenzona, \href{http://dx.doi.org/10.1016/0022-5088(81)90143-0}{Contribution to the study of the binary phase diagrams ca--pt and sr--pt}, Journal of the Less Common Metals 78~(2) (1981) 49–53.
\newblock \href {https://doi.org/10.1016/0022-5088(81)90143-0} {\path{doi:10.1016/0022-5088(81)90143-0}}.
\newline\urlprefix\url{http://dx.doi.org/10.1016/0022-5088(81)90143-0}

\bibitem{Vergniory2019}
M.~G. Vergniory, \emph{et al.}, A complete catalogue of high-quality topological materials, Nature 566 (2019) 480--485.
\newblock \href {https://doi.org/10.1038/s41586-019-0954-4} {\path{doi:10.1038/s41586-019-0954-4}}.

\bibitem{Giannozzi2009}
P.~Giannozzi, et~al., \href{https://doi.org/10.1088/0953-8984/21/39/395502}{Quantum espresso: a modular and open-source software project for quantum simulations of materials}, J. Phys.-Condes. Matter 21~(39) (2009) 395502.
\newblock \href {https://doi.org/10.1088/0953-8984/21/39/395502} {\path{doi:10.1088/0953-8984/21/39/395502}}.
\newline\urlprefix\url{https://doi.org/10.1088/0953-8984/21/39/395502}

\bibitem{Perdew96p3865}
J.~P. Perdew, K.~Burke, M.~Ernzerhof, Generalized gradient approximation made simple, Phys. Rev. Lett. 77~(18) (1996) 3865--3868.
\newblock \href {https://doi.org/10.1103/PhysRevLett.77.3865} {\path{doi:10.1103/PhysRevLett.77.3865}}.

\bibitem{hamann_optimized_2013}
D.~R. Hamann, \href{https://link.aps.org/doi/10.1103/PhysRevB.88.085117}{Optimized norm-conserving vanderbilt pseudopotentials}, Physical Review B 88~(8) (2013) 085117.
\newblock \href {https://doi.org/10.1103/PhysRevB.88.085117} {\path{doi:10.1103/PhysRevB.88.085117}}.
\newline\urlprefix\url{https://link.aps.org/doi/10.1103/PhysRevB.88.085117}

\bibitem{van_setten_pseudodojo_2018}
M.~J. van Setten, M.~Giantomassi, E.~Bousquet, M.~J. Verstraete, D.~R. Hamann, X.~Gonze, G.~M. Rignanese, \href{https://www.sciencedirect.com/science/article/pii/S0010465518300250}{The pseudodojo: Training and grading a 85 element optimized norm-conserving pseudopotential table}, Comput. Phys. Commun. 226 (2018) 39--54.
\newblock \href {https://doi.org/10.1016/j.cpc.2018.01.012} {\path{doi:10.1016/j.cpc.2018.01.012}}.
\newline\urlprefix\url{https://www.sciencedirect.com/science/article/pii/S0010465518300250}

\bibitem{monkhorst_special_1976}
H.~J. Monkhorst, J.~D. Pack, \href{https://link.aps.org/doi/10.1103/PhysRevB.13.5188}{Special points for brillouin-zone integrations}, Phys. Rev. B 13 (1976) 5188--5192.
\newblock \href {https://doi.org/10.1103/PhysRevB.13.5188} {\path{doi:10.1103/PhysRevB.13.5188}}.
\newline\urlprefix\url{https://link.aps.org/doi/10.1103/PhysRevB.13.5188}

\bibitem{Mostofi14p2309}
A.~A. Mostofi, J.~R. Yates, G.~Pizzi, Y.-S. Lee, I.~Souza, D.~Vanderbilt, N.~Marzari, An updated version of wannier90: A tool for obtaining maximally-localised wannier functions, Computer Physics Communications 185~(8) (2014) 2309--2310.

\bibitem{Damle15p1463}
A.~Damle, L.~Lin, L.~Ying, Compressed representation of kohn--sham orbitals via selected columns of the density matrix, Journal of chemical theory and computation 11~(4) (2015) 1463--1469.

\bibitem{Damle18p1392}
A.~Damle, L.~Lin, Disentanglement via entanglement: a unified method for wannier localization, Multiscale Modeling \& Simulation 16~(3) (2018) 1392--1410.

\bibitem{qeirrep}
A.~Matsugatani, S.~Ono, Y.~Nomura, H.~Watanabe, \href{https://www.sciencedirect.com/science/article/pii/S0010465521000722}{qeirreps: An open-source program for quantum espresso to compute irreducible representations of bloch wavefunctions}, Comput. Phys. Commun. 264 (2021) 107948.
\newblock \href {https://doi.org/https://doi.org/10.1016/j.cpc.2021.107948} {\path{doi:https://doi.org/10.1016/j.cpc.2021.107948}}.
\newline\urlprefix\url{https://www.sciencedirect.com/science/article/pii/S0010465521000722}

\bibitem{Wu18p405}
Q.~Wu, S.~Zhang, H.-F. Song, M.~Troyer, A.~A. Soluyanov, Wanniertools: An open-source software package for novel topological materials, Computer Physics Communications 224 (2018) 405--416.

\bibitem{Teo2008surface}
J.~C. Teo, L.~Fu, C.~Kane, Surface states and topological invariants in three-dimensional topological insulators: Application to bi 1- x sb x, Physical Review B—Condensed Matter and Materials Physics 78~(4) (2008) 045426.

\bibitem{Rauch21p195103}
T.~c.~v. Rauch, T.~Olsen, D.~Vanderbilt, I.~Souza, \href{https://link.aps.org/doi/10.1103/PhysRevB.103.195103}{Mirror chern numbers in the hybrid wannier representation}, Phys. Rev. B 103 (2021) 195103.
\newblock \href {https://doi.org/10.1103/PhysRevB.103.195103} {\path{doi:10.1103/PhysRevB.103.195103}}.
\newline\urlprefix\url{https://link.aps.org/doi/10.1103/PhysRevB.103.195103}

\bibitem{Shin2017}
D.~Shin, Y.~Lee, M.~Sasaki, Y.~H. Jeong, F.~Weickert, J.~B. Betts, H.-J. Kim, K.-S. Kim, J.~Kim, \href{http://dx.doi.org/10.1038/nmat4965}{Violation of ohm’s law in a weyl metal}, Nature Materials 16~(11) (2017) 1096–1099.
\newblock \href {https://doi.org/10.1038/nmat4965} {\path{doi:10.1038/nmat4965}}.
\newline\urlprefix\url{http://dx.doi.org/10.1038/nmat4965}

\bibitem{Jiang2022}
Y.~Jiang, M.~Ermolaev, G.~Kipshidze, S.~Moon, M.~Ozerov, D.~Smirnov, Z.~Jiang, S.~Suchalkin, \href{http://dx.doi.org/10.1038/s41467-022-33560-x}{Giant g-factors and fully spin-polarized states in metamorphic short-period inassb/insb superlattices}, Nature Communications 13~(1) (Oct. 2022).
\newblock \href {https://doi.org/10.1038/s41467-022-33560-x} {\path{doi:10.1038/s41467-022-33560-x}}.
\newline\urlprefix\url{http://dx.doi.org/10.1038/s41467-022-33560-x}

\end{thebibliography}

\end{document}